# Tunable Multistage Refrigeration via Geometrically Frustrated Triangular Lattice Antiferromagnet for Space Cooling

Jianqiao Wang[a,b,§], Chushu Fang[b,c,§], Zhibin Qiu[a,b], Yang Zhao[a,b,d], Quan Xiao[a,b,d], Xiying Sun[a,b], Zhaoyi Li[a,b,d], Laifeng Li[c], Yuan Zhou[c], Changzhao Pan[b,]*, Shu Guo[a,b,]*

[a]Shenzhen Institute for Quantum Science and Engineering, Southern University of Science and Technology, Shenzhen 518055, China

[b]International Quantum Academy, Shenzhen 518048, China

[c]State Key Laboratory of Cryogenics, Technical Institute of Physics and Chemistry, Chinese Academy of Sciences, Beijing 100190, China

[d]School of Physics and Electronics, Henan University, Kaifeng 475004, China

**Abstract**

Low-temperature refrigeration technology constitutes a crucial component in space exploration. The small-scale, low-vibration Stirling-type pulse tube refrigerators hold significant application potential for space cooling. However, the efficient operation of current Stirling-type pulse tube cryocoolers in space cooling applications remains challenging due to the rapid decay of the heat capacity of regenerative materials below 10 K. This study adopts a novel material strategy: using a novel high-spin $S = 7/2$ magnetic regenerative material, $Gd_2O_2Se$, we construct a multistage tunable regenerative material structure to achieve an efficient cooling approach to the liquid helium temperature range. Under substantial geometric frustration from a double-layered triangular lattice, it exhibits two-step specific heat transition peaks at 6.22 K and 2.11 K, respectively. Its ultrahigh specific heat and broad two-step transition temperature range effectively bridge the gap between commercially used high-heat-capacity materials. Experimental verification shows that when $Gd_2O_2Se$ is combined with $Er_3Ni$ and $HoCu_2$ in the Stirling-type pulse tube cryocooler, the cooling efficiency of the pulse tube increases by 66.5 % at 7 K, and the minimum achievable temperature reaches 5.85 K. These results indicate that $Gd_2O_2Se$ is an ideal magnetic regenerative material for space cooling.

**Keywords:** Magnetic regenerative material; Geometrical frustration; Stirling type pulse tube refrigerator; Space cooling

## 1. Introduction

Deep space exploration technology, a crucial component of cosmic observation, has an intense demand for space cooling technologies[1-3]. For example, the James Webb Space Telescope, released by NASA in 2002, requires a maintained operating temperature of 6 K to ensure the stable operation of its onboard Mid-Infrared Instrument[2]. Furthermore, projects such as the China Space Station Telescope and the Hot Universe Baryon Surveyor satellite program also have high demands for space cryogenic cooling technologies[4-6]. Meanwhile, the Stirling pulse tube refrigerators (SPTRs) have gradually emerged as a research focus in the development of space cooling technologies, owing to their advantages of having no mechanical components in the cold head, long lifespan, low vibration, and light weight[7-9]. However, their efficiency below 10 K temperatures for space cooling applications remains suboptimal. A primary reason for this is that the heat capacity of commonly used magnetic regenerative materials decreases as the temperature drops. Therefore, the search for materials that can maintain high-heat-capacity near liquid helium temperature operating ranges is crucial for the development of cryogenic coolers for applications in space cooling[10].

In the early days, lead (Pb), as a commonly used regenerative material, exhibited high volumetric specific heat at low temperatures ($C_p$ = 0.35 J·K$^{-1}$·cm$^{-3}$ at 15 K)[11]. However, the specific heat of lead is primarily contributed by lattice vibrations; when the temperature drops below 10 K ($T \ll \Theta_D$), its phonon degrees of freedom decrease, causing the specific heat to approach zero gradually and the cryocooler to fail[12]. The systematic development of rare-earth-based (*RE*-based) regenerative materials began in the late 1980s and early 1990s, making it possible for reciprocal cryocoolers to reach liquid helium temperatures[13]. By leveraging spin ordering transitions in magnets during cooling, high specific heat at low temperatures is achieved, thereby ensuring the performance of reciprocal cryocoolers in cryogenic regimes[14,15]. Over decades of development, *RE*-based alloys and oxides, such as Gd$_3$Ga$_5$O$_{12}$(GGG), GdAlO$_3$, HoCu$_2$, Er$_3$Ni, and Gd$_2$O$_2$S (GOS), have been successively developed and applied to refrigeration in the liquid helium temperature range[16-19]. In the application of SPTR, by utilizing the relay of material transition temperatures, multi-stage cooling regulation within the SPTR can be achieved through two-end or even multi-section filling of regenerative materials. For example, in 2018, Chen et al. investigated a 4 K gas-coupled SPTR precooled by liquid nitrogen, achieving a no-load temperature of 3.6 K through a 38% $HoCu_2$ and 62% $Er_3Ni$ filling[20]. In 2020, Dang et al. developed a four-stage SPTR, using 60% $Er_3Ni$ and 40% $HoCu_2$ as regenerative materials for the fourth stage, which reached a no-load temperature of 4.2 K[21]. In 2021, Qiu et al. employed GOS and $HoCu_2$ as filling materials for the SPTR, achieving a no-load temperature of 4.57 K[22]. However, in previous studies, the cooling performance of the refrigerator filled with GOS deteriorates when the temperature exceeds 6.1 K. The main reason is that there is a heat-capacity gap between GOS and $HoCu_2$ in the temperature range between 5.2~7.3 K, which motivates us to address this gap through materials science approaches.

Among numerous *RE*-based compounds, $Gd^{3+}$- or $Eu^{2+}$-containing materials are widely studied due to their zero orbital angular momentum, along with the maximum spin angular momentum ($S$ = 7/2) and the highest magnetic entropy per ion (Rln8) [23-25]. This unique advantage of single-ion magnetism also renders the controlled spatial assembly of $Gd^{3+}$ spin centers an effective means to regulate refrigerant performance. On the one hand, by adjusting the ratio of $Gd^{3+}$ magnetic ions to ligands, the magnitude of magnetic entropy can be tuned to achieve ideal performance in multistage SPTR filling; on the other hand, regulation of magnetic ion density influences magnetic interactions, thereby enabling control over the magnetic transition temperature in the *B* = 0 T ground state and achieving the relay of specific heat transition temperatures for multistage SPTR filling materials[26-28]. In Gd-based complexes, GOS exhibits pronounced antiferromagnetic interactions and a magnetic ion density as high as 83%, along with ultra-high volumetric specific heat at 5.2 K and excellent mechanical properties[18]. Since its first report in 2003, GOS has evolved into a widely used commercial regenerative material[29]. As the isostructural for GOS, $Gd_2O_2Se$ (GOSe), was synthesized via isovalent element substitution in GOS, it maintains the same high-symmetry structure and a high magnetic ion density of 74%. Past studies have demonstrated that it possesses excellent optical properties, similar to those of GOS[30,31].

It is expected that it can form a relay within the transition temperature gap of existing commercial refrigeration materials, construct a multistage SPTR, and achieve efficient refrigeration in space cooling and the liquid helium temperature range. Moreover, due to the stringent requirements for the vacuum degree in the synthesis process, the research on GOSe has mostly focused on micrometer-scale dimensions, which has restricted its large-scale application.

In this work, refrigeration performance and magnetic specific heat capacity have been investigated in the novel magnetic regenerative material GOSe. Structurally, the synthesized millimeter-scale GOSe single crystal has a $P$-$3m1$ space group, thus arranging $Gd^{3+}$ magnetic ions into an equilateral triangular lattice (TL) pattern, which is favorable for geometrically magnetic frustration and strong magnetic exchange interactions. As a result, the GOSe exhibits two-step specific heat transition peaks at 6.22 K and 2.11 K, respectively, featuring an ultrahigh specific heat peak value of 1.02 J·cm$^{-3}$·K$^{-1}$ and a broad operating temperature range for regenerative cooling applications. After being filled into the pulse tube together with $Er_3Ni$ and $HoCu_2$, the refrigeration capacity of GOSe in the range of 4 - 35 K has been verified. It is worth noting that in the SPTR, the cooling efficiency of the pulse tube increases by 66.5 % at 7 K, and the minimum achievable temperature reaches 5.85 K. Our results show that GOSe exhibits excellent performance among commercial regenerative materials, providing a novel solution for lightweight and high-efficiency space refrigeration systems.

## 2. Results and discussion

### 2.1. Crystal structure.

The process of preparing GOSe single crystals using the chemical vapor transport method is shown in **Figure 1a**, and it is described in the experimental section. The crystal structure of GOSe was examined by single-crystal X-ray diffraction (SC-XRD). As expected, the space group of GOSe at 300 K is *P*-3*m*1 ($a = b = 3.93(2)$ Å and $c = 6.92(4)$ Å). **Figure 1b** shows the as-grown GOSe crystals, with a perfect plate-like structure, exhibiting macroscopic trigonal crystal symmetry with millimeter-scale *ab*-plane dimensions of $0.96 \times 0.88$ mm$^2$. Combining with the crystal structures of GOSe shown in **Figures 1a**, **c**, and **d**, this perfect trigonal crystal structure originates from the equilateral-triangle arrangement of Gd$^{3+}$ ions. Each Gd atom is coordinated by four O and three Se atoms, forming a seven-coordinate polyhedron described as a one-capped distorted trigonal antiprism (**Figure 1c**)[30]. The prismatic structures composed of four oxygen atoms and Gd atoms are alternately arranged, forming a Gd-based double-layered TL, which may lead to the geometric frustration effect and generate strong magnetic interactions (**Figures 1d and 1e**)[32,33]. **Figures 1f–h** show the reciprocal planes (hk0), (h0l), and (0kl) from SC-XRD data collected at room temperature. Notably, on the left half of each panel, we simulated the diffractions of the GOSe lattice using Single Crystal software[34], which are consistent with the experimental diffraction patterns. The absence of twin diffraction spots indicates a good crystalline quality. Rietveld refinement of the powder X-ray diffraction (P-XRD) data shows that all observed peaks match the calculated pattern and correspond to the expected Bragg positions (**Figure 1i**), confirming the phase purity and validating the structural model.

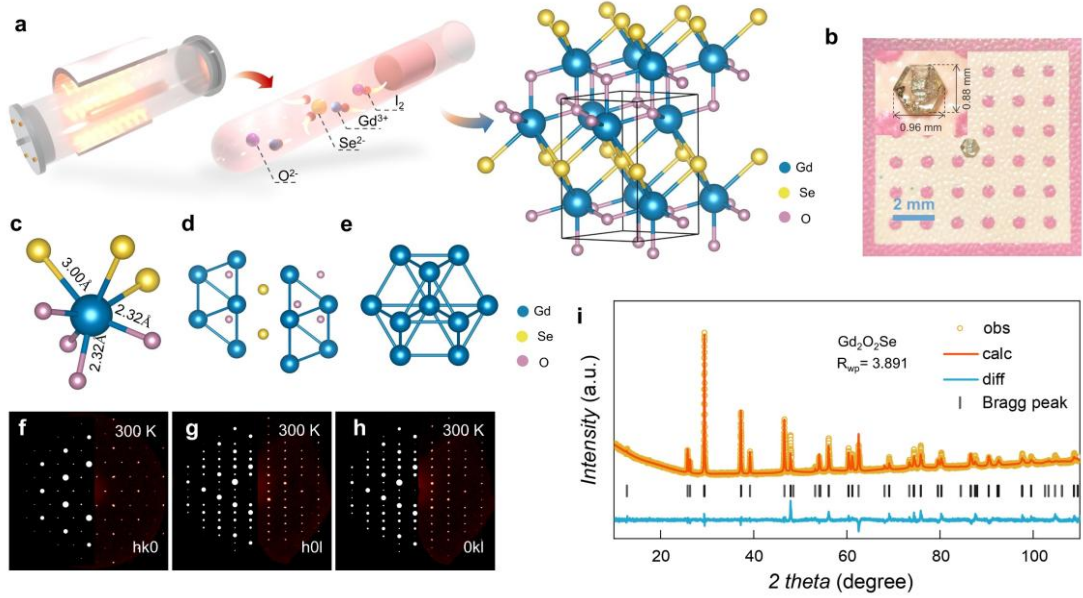

**Figure 1. Synthesis, structure, and phase purity.** (a) Schematic diagrams of the GOSe's synthesis method and its crystal structure, (b) Crystal picture of GOSe. (c) The coordination environment of $Gd^{3+}$ ions in the GOSe crystal, as well as the ionic arrangement structures of $Gd^{3+}$ ions within the (d) *a-b* plane and (e) along the *c*-axis. Reciprocal lattice planes (f) (*hk*0), (g) (*h*0*l*), (h) (*0kl*) from SC-XRD data. (i) Rietveld refinement of the P-XRD pattern measured at room temperature.

## 2.2. Magnetism of GOSe.

**Figures 2a** and **b** show the temperature-dependent magnetic susceptibility of GOSe under a 0.05 T magnetic field applied along different directions. In the high-temperature region, the $\chi(T)$ curves exhibit typical Curie-Weiss (C-W) behavior. The zero-field cooling (ZFC) and field cooling (FC) curves in two different orientations coincide well, and no bifurcations between the ZFC and FC curves commonly seen in spin glass systems are observed[35,36]. And it can be seen from **Figure 2a** that due to the long-range ordering of the spin directions of magnetic $Gd^{3+}$ ions, an antiferromagnetic (AFM) transition occurs around 6.39 K, accompanied by an additional anomaly near 2 K under a small magnetic field of 500 Oe. When the same magnetic field is applied along the *c*-axis (**Figure 2b**), the magnetic susceptibilities of the crystal exhibit slight differences, specifically manifested as two clear magnetic transitions appearing around 6.29 K (AFM1) and 2.17 K (AFM2), respectively. To further probe the magnetic behavior of GOSe, the magnetic susceptibility along two directions was measured under external magnetic fields of 0.05 - 5 T, as shown in **Figures 2c** and **d**. For $B \perp c$, below the AFM1, magnetic susceptibility exhibits an increase with applied magnetic field (**Figure 2c**). Under weak fields ($B < 2$ T), the higher-temperature AFM1 at $T_{N1}$ shifts slightly to lower temperatures and broadens, while the lower-temperature transition AFM2 at $T_{N2}$ is gradually suppressed and becomes indiscernible at $B = 2$ T. This phenomenon likely originates from the paramagnetic-to-AFM1 transition at $T_{N1}$ under low magnetic fields, followed by the emergence of a lower-energy AFM2 state at $T_{N2}$ driven by strong geometric frustration. This phenomenon of two inflection points also occurs in the $B \parallel c$ direction, and repeated tests on different samples all yield such results (**Figure 2d**). Given the nearly isotropic magnetic behavior for GOSe, $1/\chi(T)$ curve of the GOSe single crystal under $B = 0.5$ T with $B \perp c$ was selected as a representative for further analysis. (**Figure 2e**). When the temperature is much higher than the Néel temperature, the $1/\chi(T)$ curve exhibits linear behavior. The C-W law can be expressed as[37]:

$$\chi = \frac{C}{T - \Theta_{cw}} + \chi_0 \quad (1)$$

Where the $C$ and $\Theta_{cw}$ represent the Curie constant and the paramagnetic C-W temperature, respectively. and $\chi_0$ is the temperature-independent magnetic susceptibility contribution. The negative C-W temperature ($\Theta_{cw} = -7.76$ K) of GOSe confirms the dominant antiferromagnetic interactions. And the $\mu_{eff}$ is calculated using the $\mu_{eff} = \sqrt{8C}\mu_B$. The $\mu_{eff}$ calculated based on the magnetic fields $B \perp c$ is 7.54 $\mu_B$, which is a little smaller than the $Gd^{3+}$ magnetic moment (7.9 $\mu_B$) predicted by Hund's rules[38]. The $\mu_{eff}$ calculated based on the magnetic fields $B \parallel c$ is 7.53 $\mu_B$, which is close to that obtained under $B \perp c$ (**Figure S4**). **Figure 2f** displays the field-dependent magnetization for $B \parallel c$ and $B \perp c$ at different temperatures between 1.8 and 10 K. At high temperature (10 K), the magnetization exhibits behavior close to linearity. At 1.8 K, the field-dependent magnetization shows nonlinear behavior and does not saturate at $B = 14$ T. Along the $B \perp c$ direction, the maximum magnetization measured at 14 T, $M_{s\parallel} = 6.91$ $\mu_B/Gd^{3+}$, is close to the expected saturation moment of the system (~7.0 $\mu_B$ per $Gd^{3+}$). The behavior along the $B \parallel c$ direction is consistent with that of the $B \perp c$, with a maximum magnetization $M_{s\perp}$ = 6.69 $\mu_B$/Gd$^{3+}$, approaching isotropic magnetic behavior. This isotropy enables GOSe to be used in SPTR and other cryocoolers without considering crystal orientation, achieving magnetic regenerative cooling with high efficiency. The magnetic properties of ground GOSe powder (**Figure S5**) show consistent susceptibility and magnetization for both $B // c$ and $B \perp c$ directions, indicating the convenient usability of GOSe for magnetic refrigeration.

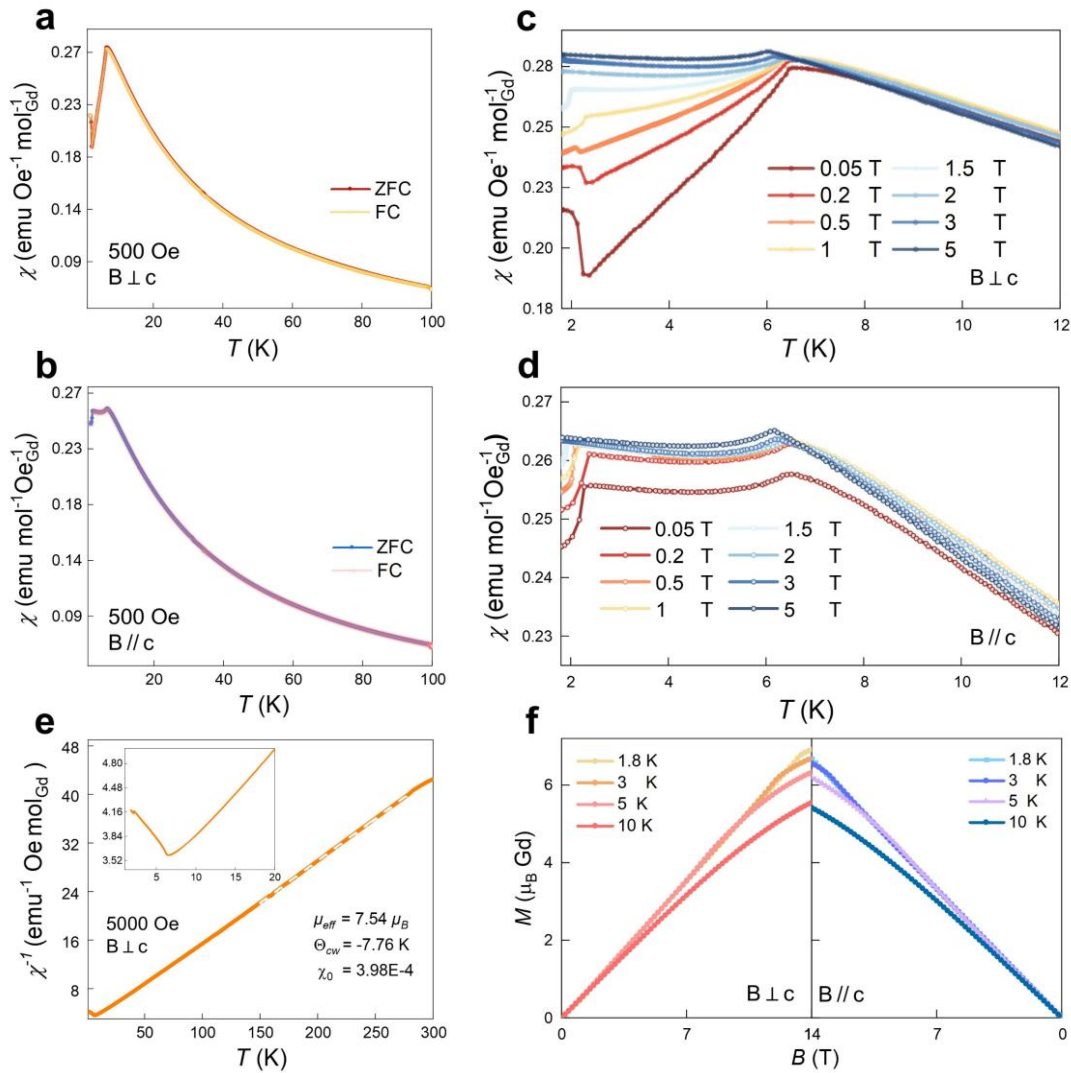

**Figure 2. Magnetism.** Temperature-dependent magnetic susceptibility measured under a magnetic field of 0.05 T for (a) $B \perp c$ and (b) $B // c$. Field-dependence of magnetic susceptibility at different temperatures for (c) $B \perp c$ and (d) $B // c$. Inverse magnetic susceptibility as a function of temperature under a magnetic field of 0.5 T for (e) $B \perp c$. Field-dependent magnetization at selected temperatures for $B \perp c$ and $B // c$ (f).

**2.3. Magnetic regenerative properties of GOSe.**

To elucidate the specific heat characteristics of the GOSe crystal, the specific heat were performed under different magnetic fields ($B // c$, 0–14 T), as shown in **Figure 3a.** The zero-field specific heat curve exhibits a $\lambda$ peak around 6.22 K, indicating the onset of a second-order AFM1 transition, which is in good agreement with the temperature-dependent magnetic susceptibility results. Upon applying a magnetic field, spins tend to align along the field, weakening antiferromagnetic correlations and causing the specific heat peak to shift gradually to lower temperatures while its magnitude diminishes. Even at 14 T, the specific-heat anomaly is not fully suppressed—only a weak, broadened peak remains and shifts in temperature—consistent with the lack of magnetization saturation at 14 T. These observations suggest that AFM correlations persist to high fields, indicative of robust exchange interactions. At zero field, the specific heat peak reaches 55.8 J·mol$^{-1}$·K$^{-1}$ at 6.22 K, and magnetocaloric materials operate near liquid helium temperatures.

In the higher temperature region, the lattice contribution arising from thermal vibrations dominates the heat capacity, which can be subtracted via modified Debye mode fitting[39]:

$$C_{p,ph} = 9R \sum_{n=1}^{2} C_n \left(\frac{T}{\theta_{Dn}}\right)^3 \int_0^{\frac{\theta_{Dn}}{T}} \left(\frac{x^4 e^x}{(e^x - 1)^2}\right) dx \qquad (2)$$

where the sum of $C_1$ plus $C_2$ should be equal to 1, and the fitting interval is selected as 25–50 K. Based on the fitting results, the phonon specific heat extends to 1.8 K (**Figure S7**). By subtracting the phonon specific heat $C_{p,ph}$ from the total specific heat $C_p$, the magnetic specific heat $C_{p,mag}$ is obtained. The magnetic entropy with different magnetic fields is calculated via integration from 1.8 to 50 K by[39]:

$$S_{mag} = \int \left(\frac{C_{mag}}{T}\right) dT \qquad (3)$$

The magnetic entropy reaches to 13.74 J·mol$^{-1}$·K$^{-1}$ at 50 K under zero field, which is lower than the expected value for a spin-7/2 system (17.29 J·mol$^{-1}$·K$^{-1}$), suggesting additional magnetic entropy remains to be released at lower temperatures (**Figure 3b**). Furthermore, magnetic entropy is gradually suppressed with increasing magnetic field, which is consistent with the trend described for the specific heat.

To further assess the magnetic entropy below 1.8 K, specific heat measurements using a $^3$He setup reveal a sharp, narrow peak at 2.11 K in GOSe (**Figure 3c**), which is characteristic of a first-order transition and is consistent with the magnetic susceptibility results (**Figure 2d**). The magnetic entropy was further evaluated to be 16.88 J·mol$^{-1}$·K$^{-1}$, consistent with the expected value for an $S = 7/2$ state in GOSe. As shown in **Figure 3d,** with the gradual increase of the magnetic field, the transition peak shifts progressively toward lower temperatures, corresponding to the AFM2 transition. Such two-step AFM transitions enable a broad operating temperature range for specific heat applications. The AFM2 anomaly around 2.11 K is not clearly observed in the $^4$He specific heat measurements, likely because it is smeared out or masked by background contributions close to the lower temperature limit (~1.8 K) of the $^4$He platform.

For comparison, the specific heat performance of GOSe with that of typical regenerative materials operating near liquid helium temperatures is presented in **Figure 3e** [40]. Notably, the specific heat of GOSe perfectly fills the gap between two commonly used regenerative materials, GOS and HoCu$_2$, while exhibiting an excellent transition peak that is higher than that of commercial materials such as Er$_3$Ni and Er$_{0.9}$Yb$_{0.1}$Ni[16,17,41,42]. Furthermore, the broad two-step specific heat transition peaks exhibited by GOSe have significantly expanded the regenerative cooling capacity and application range of the material.

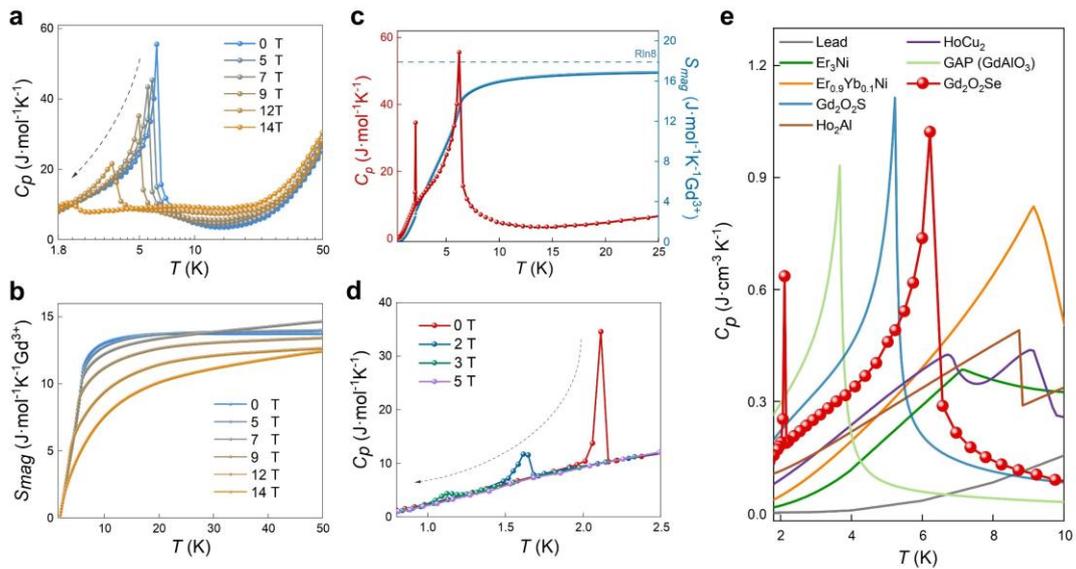

**Figure 3. Thermodynamic Properties of GOSe.** (a) Temperature-dependent specific heat of GOSe.

(b) Temperature-dependent magnetic entropy of GOSe. (c) Specific heat and magnetic entropy of GOSe measured from 0.4 to 50 K under zero field. (d) The variation of the low-temperature specific heat transition peak of GOSe with 0 - 5 T magnetic field. (e) The zero-field specific heat of GOSe is compared with the commercially available regenerative materials.

## 2.4. Cooling performance of SPTR based on GOSe.

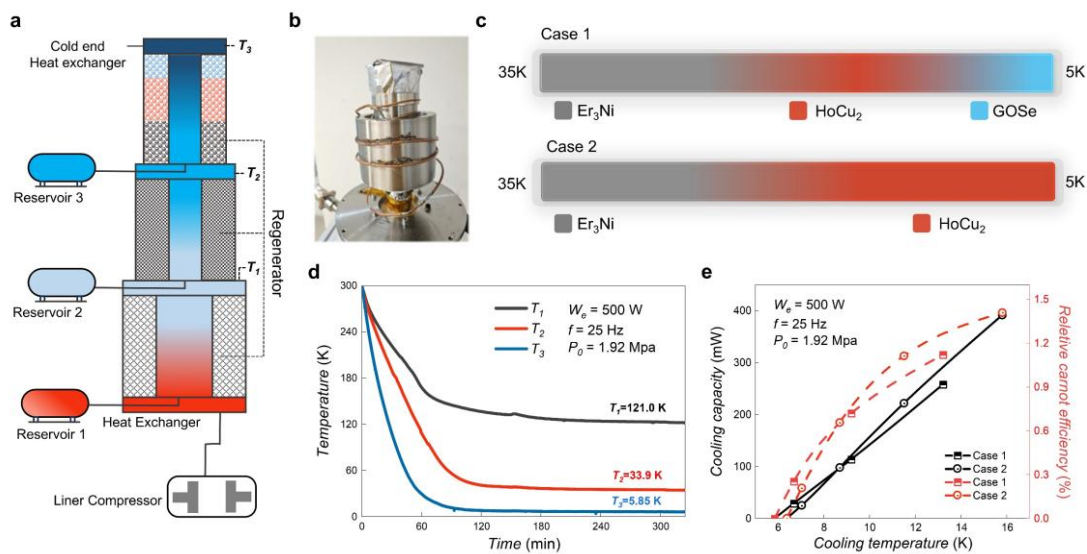

**Figure 4. The cooling performance of the SPTR cryocooler.** (a) Schematic diagram of the cryocooler structure. (b) Physical drawing of a three-stage high-frequency pulse tube cryocooler. (c) Comparative illustrations of filling cases for cryocooler regenerators' low-temperature-stage, case 1: filled with GOSe; case 2: without GOSe. (d) Cooling curve of the cryocooler after using GOSe. (e) Cooling capacity and relative Carnot efficiency of case 1 and case 2.

According to the existing three-stage SPTR model, we investigated the application of GOSe materials in a three-stage high-frequency pulse tube cryocooler[43]. As shown in **Figure 4a,** the three-stage system consists of a first-stage regenerator, a first-stage precooling heat exchanger, a second-stage regenerator, a second-stage precooling heat exchanger, a third-stage regenerator, a cold-end heat exchanger, a pulse tube, and a reservoir. The corresponding physical diagram of the cryocooler is shown in **Figure 4b**. This cryocooler previously achieved an experimental minimum refrigeration temperature of 6.38 K using $HoCu_2$ and $Er_3Ni$ as filling materials. **Figure 4c** illustrates the packing configuration of GOSe in the regenerator of the low-temperature stage of the cryocooler. In Case 1, the regenerator is divided into three temperature zones corresponding to the specific heat peak positions of different regenerative materials, with the packing sequence as follows: the coldest end is filled with GOSe, the middle section with $HoCu_2$, and the hot end with $Er_3Ni$. For comparison, Case 2 shows the packing configuration adopted by the regenerator of the low-temperature stage in the reference cryocooler.

The filling amount of the material was determined via simulation. As shown in **Figure S8**, four cases with different lengths of GOSe are simulated. The relationship between the cooling power ($Qc$) and the refrigeration temperature ($T_{base}$) is shown in **Figure S9a**. Compared to the case without GOSe, the no-load refrigeration temperatures of cases with GOS at the cold end of the regenerator decrease from 3.2 K to a minimum of 3.05 K. The cooling power could be increased by GOSe at low temperatures. However, when the no-load refrigeration temperature is above 7 K, the GOSe spheres would no longer help improve the cooling performance. **Figure S9b** shows the relationship between $Qc$ and the length of GOSe. As can be seen, the lower the $T_{base}$ is, the longer the optimal length of GOSe will be. Specifically, at a $T_{base}$ of 5 K, filling 9 - 10 mm of GOSe brings out an optimal cooling power of approximately 35 mW, which is 6 mW higher than that in the absence of GOSe.

In the three-stage high-frequency pulse tube cryocooler, the compressor operates with 500 W input power at 25 Hz frequency, achieving a stabilized front-chamber pressure of 1.92 MPa following extended cooling. **Figure 4d** presents the three-stage cooling curves recorded after 5.5 hours of operation, with final cold end temperatures reaching 5.85 K (third stage), 33.9 K (second stage), and 121.0 K (first stage). Comparative experiments were performed on two regenerator filler configurations (case 1: with GOSe; case 2: without GOSe) shown in **Figure 4c,** for a high-frequency pulse tube cryocooler. As shown in **Figure 4e**, the GOSe-enhanced cryocooler (case 1) demonstrates performance improvement down to 8 K. At 7 K, case a achieves 39.3 mW refrigeration capacity with 0.334% relative Carnot efficiency, showing 66.5% enhancement over case 2's 23.6 mW at 0.198% efficiency. Notably, the GOSe advantage diminishes at higher temperatures, with case b exhibiting better cooling capacity (specified in **Table 1**) at 10 K. Detailed configuration parameters and performance metrics are systematically compared in **Table 1**.

It is worth noting that compound GOSe, together with its parent commercial material GOS, offers opportunities for systematic tuning of material properties through substitutions at both magnetic and nonmagnetic sites. More notably, the prototype refrigerator incorporating GOSe has an overall mass not exceeding 20 kg, exhibiting a distinct lightweight advantage for space cooling applications. Consequently, our preliminary findings not only shed light on the space cooling capacity of GOSe but also point to new directions for developing efficient cryogenic refrigeration technologies in the liquid-helium temperature regime, enabled by this unique frustrated magnet.

**Table 1.** Specific filling methods and cooling performance of the two filling cases.

| Case | Filling method | Base Temperature (K) | Cooling capacity (500 W electric power) |
|---|---|---|---|
| 1 | With GOSe (GOSe+ HoCu$_2$+ Er$_3$Ni) | 5.85 | 39.3 mW/7 K |
| 2 | Without GOSe (HoCu$_2$+ Er$_3$Ni) | 6.38 | 23.6 mW/7 K |

## 3. Conclusion

In this work, excellent refrigeration performance and magnetic specific heat capacity have been investigated in the novel geometrically frustrated magnet GOSe. In terms of structure, the high symmetry space group ($P$-3$m$1) arranges the Gd$^{3+}$ magnetic ions in an equilateral TL pattern. As a result, the GOSe exhibits two-step specific heat transition peaks around 6.22 K and 2.11 K, featuring an ultrahigh specific heat peak value of 1.02 J·cm$^{-3}$·K$^{-1}$ and a broad operating temperature range for regenerative cooling applications. After being filled into the pulse tube together with GOS and HoCu$_2$, the refrigeration capacity of GOSe in the range of 4 - 35 K has been verified. It is worth noting that in the Stirling-type pulse tube cryocooler, the cooling efficiency of the pulse tube increases by 66.5 % at 7 K, and the minimum achievable temperature reaches 5.85 K, demonstrating excellent performance among regenerative materials. Our results demonstrated that GOSe is an excellent magnetic regenerative material for space cooling.

## 4. Method

**Single crystal growth**: Referring to past literature[30,44], the single crystals of GOSe were prepared via the reaction:

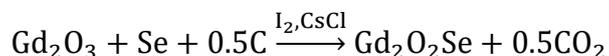

$$Gd_2O_3 + Se + 0.5C \xrightarrow{I_2, CsCl} Gd_2O_2Se + 0.5CO_2$$

High-purity $Gd_2O_3$, Se, and C were weighed and uniformly ground in an agate mortar at a molar ratio of 2:4:1. Trace amounts of $I_2$ and CsCl (three times the molar amount of $Gd_2O_3$) were added to a quartz ampoule, which was then evacuated and sealed. The ampoule was heated to 1000 °C in a muffle furnace and held for 48 hours, followed by slow cooling to 700 °C inside the furnace and further gradual cooling to room temperature. The product was washed with distilled water to remove cesium chloride and dried in air, yielding transparent hexagonal single crystals.

**P-XRD:** High-quality single crystals were selected under an optical microscope for thorough grinding. The obtained powder sample was then spread flat on a special glass slide and installed on the sample stage of the Rigaku SmartLab 9 KW instrument. The Bragg-Brentano geometric optical path was chosen for diffraction experiments. The light source used Cu K$\alpha$ (K$\alpha$1 = 1.54056 Å, K$\alpha$2 = 1.54439 Å) radiation. The program settings were as follows: the scanning range was 10°~110°, the step length was 0.02°, and the scanning speed was 1.5°/min. Finally, the experimental data were refined using GSAS II software[45]. And use the crystal structure parameters obtained from SC-XRD, Rietveld refinement of the P-XRD patterns was performed using the GSAS II software package. $R_{wp}$, representing the weighted profile variance of the Rietveld refinement, The $R_{wp}$ = 3.89, which confirms that the phase of the bulk powdered GOSe is consistent with that of the pure single crystal.

**SC-XRD:** SC-XRD experiments were conducted on a Bruker D8 VENTURE diffractometer equipped with a PHOTON III CPAD detector. The experiments were performed at both 300 K and 100 K in a temperature-controlled system utilizing nitrogen ($N_2$). X-rays ($\lambda$ = 0.71073 Å) were generated using a graphite monochromated Mo target. Background, polarization, and Lorentz factor corrections were subsequently applied using the APEX4 software, and multi-scan absorption correction was carried out with the SADABS package. The direct method of the ShelXT program was used for the structural solution, and the ShelXL least squares refinement package of Olex2 software was used for structural refinement[46,47]. The crystal structure was subjected to X-ray simulation using the Single Crystal 5 software package, yielding weighted diffraction points with structure factors[34,48]. In **Figure S1**, the diffraction planes collected at 100 K are shown, with the relevant crystallographic data listed together in **Table S1**. As a result, no structural phase transition occurred during the cooling process from 300 K to 100 K. The lattice parameters exhibited slight contraction due to the temperature decrease, indicating the structural stability of the crystal.

**Elemental analysis:** Elemental analysis was performed using scanning electron microscopy (SEM) and energy-dispersive X-ray spectroscopy (EDS). As shown in **Figure S2**, Gd and Se elements are uniformly distributed on the trigonal crystals and exhibit a 2:1 ratio, as shown in **Figure S3**, confirming the successful formation of GOSe.

**Magnetic measurements:** Direct-current (DC) magnetic susceptibility was measured using a Physical Property Measurement System (PPMS) developed by Quantum Design, utilizing the Vibrating Sample Magnetometer (VSM) option of the PPMS. Based on a well-shaped single crystal piece (with a mass of 3 mg), the magnetic field direction was adjusted to be parallel to the TL layer (in-plane magnetic field or $B \perp c$) and perpendicular to the TL layer (out-of-plane magnetic field or $B//c$) for testing, as shown in **Figure S6a** and **6b.** The powder sample is measured by compacting 3 mg of the pure GOSe powder and fixing it in a copper rod with a capsule.

**Specific heat measurements:** The specific heat measurement was conducted using the adiabatic heat pulse method. Specific heat of the crystals was measured on the PPMS with different vertically applied magnetic fields ($B // c$), as depicted in **Figure S6c**. The sample was composed of 5 pieces of pure GOSe single crystals, with a mass of 1.34 mg. The temperature range for the measurement was 1.8 K $\leqslant T \leqslant$ 100 K. Furthermore, after integrating the $^3$He module, the lower temperature specific heat measurements (0.4 - 3 K) were performed on a 0.5 mg GOSe sample with $B //c$.

**Simulation method:** The simulation results were obtained using SAGE, a system-level computational tool developed by Gedeon Associates in the United States[49,50]. Originally applied to Stirling-type refrigerators, the software has evolved through multiple versions and can now simulate system-level performance for various refrigeration systems operating at cryogenic temperatures. This software enables component-level modeling of refrigerator subsystems with configurable parameters. These components are interconnected via mass and heat flows, yielding a one-dimensional (1D) steady-state solution. A pertinent computational model has been established in prior research[43]. The specific structural parameters are provided in **Table S2**. Four cases with different lengths of $Er_3Ni$, $HoCu_2$, and GOSe are simulated. The $Er_3Ni$, GOSe, and $HoCu_2$ sphere diameters are 0.07 mm, and the porosity of the matrices is 0.4. The cold end temperature varies from 4.2 to 7.5 K. GOSe is filled at the cold end of the third-section regenerator. The operating frequency is 25 Hz, the charge pressure is 2.0 MPa, and the pressure ratio at the hot end is 1.38 in the simulation.

**Cooling performance measurements:** The compressor employed in the three-stage Stirling pulse tube cryocooler (PTC) is a CP5570 moving-iron linear compressor manufactured by Lihan Cryogenics (LC). The entire system is driven by this single compressor. The experimental investigation focused specifically on the cooling performance of the third-stage regenerator within the 5 K to 35 K temperature range. Temperature measurement points include the first-stage cold-end heat exchanger ($T_1$), the second-stage cold-end heat exchanger ($T_2$), and the lowest-temperature stage ($T_3$). The $T_3$ sensor was a Lakeshore-calibrated Cernox™ temperature sensor with an accuracy of ±0.1 K. The other two sensors ($T_1$ and $T_2$) were PT100 platinum resistance thermometers. Both the PT100 and the Cernox™ sensor utilized a four-wire configuration to eliminate lead resistance effects. To minimize conductive heat leaks via the sensor leads, all wires had a diameter of 0.1 mm and were wound around the outer wall of the regenerator to utilize its cooling capacity for pre-cooling. The excitation current for the PT100 sensors (1 mA) was supplied by an NI 9217 acquisition card. Due to the lower excitation current requirement (10 µA) of the Cernox™ sensor, it was measured separately using a Lakeshore Model 218 temperature monitor. A custom temperature monitoring program, developed in LabVIEW, enabled real-time observation of the cooldown curves at each stage, facilitating the analysis of issues arising during experiments. A resistive heater, constructed from nichrome wire (140 Ω/m), was installed on the third-stage cold head. To minimize conductive heat loss through the heater leads, the wire was wound onto the sidewall of the cold head. The heater power supply was a Tektronix PWS2326 DC power source (single channel, 0–32 V, 6 A max), offering a voltage accuracy of 1 mV and a current accuracy of 1 mA. Upon reaching thermal equilibrium at the cold head, the electrical power supplied to the heater can be equated to the net refrigeration power at that stage.

**Supporting Information**

Supporting Information is available from the Wiley Online Library or from the author.

[CCDC numbers 2483906 and 248390 contain the supplementary crystallographic data for GOSe under 100 K and 300 K, respectively. These data can be obtained free of charge from The Cambridge Crystallographic Data Centre via www.ccdc.cam.ac.uk/data_request/cif.]

**Acknowledgments**

The authors acknowledge the financial support from the National Natural Science Foundation of China (22205091), the Guangdong Pearl River Talent Plan (2023QN10C793, 2023QN10X795), the National Key R&D Program of China (Grant No. 2023YFF0721303), and the Shenzhen Science and Technology Program (Grant No. KJZD20231023092601003). The authors also thank the support from the Hefei National Laboratory.

J. Q. Wang and C. S. Fang contributed equally to this work.

**Conflict of Interest**

The authors declare no conflict of interest.

**Data Availability Statement**

The data that support the findings of this study are available from the corresponding author upon reasonable request.

**TOC**

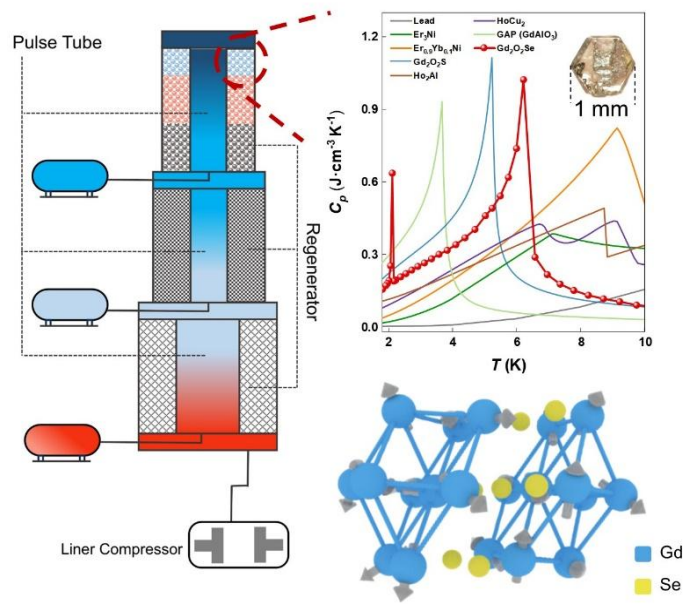

A novel geometrically frustrated magnet, $Gd_2O_2Se$, featuring a $Gd^{3+}$-based double-layered triangular lattice, exhibits two-step transitions, ultrahigh magnetic specific heat, and a broad operational range. Integrated into a multistage pulse tube cryocooler, A enhances cooling efficiency by 66.5% at 7 K and lowers the minimum temperature to 5.85 K, enabling advanced space cooling.